# CHECKERBOARD PROBLEM TO TOPOLOGY OPTIMIZATION
# OF CONTINUUM STRUCTURES

## Jun-ichi Koga, Jiro Koga and Shunji Homma

Division of Materials Science, Graduate School of Science and Engineering, Saitama University,
Saitama, 338-8570 Japan

**Abstract**: The area of topology optimization of continuum structures of which is allowed to change in order to improve the performance is now dominated by methods that employ the material distribution concept. The typical methods of the topology optimization based on the structural optimization of two phase composites are the so-called variable density ones, like the SIMP (Solid Isotropic Material with Penalization) and the BESO (Bi-directional Evolutional Structure Optimization). The topology optimization problem refers to the saddle-point variation one as well as the so-called Stokes flow problem of the compressive fluid. The checkerboard patterns often appear in the results computed by the SIMP and the BESO in which the $Q_1$-$P_0$ element is used for FEM (Finite Element Method), since these patterns are more favourable than uniform density regions. Computational experiments of SIMP and BESO have shown that filtering of sensitivity information of the optimization problem is a highly efficient way that the checkerboard patterns disappeared and to ensure mesh-independency. SIn this paper, we discuss the theoretical basis for the filtering method of the SIMP and the BESO and as a result, the filtering method can be understood by the theorem of partition of unity and the convolution operator of low-pass filter.

### Introduction

A common structural optimization in mechanical structural design field is three categorized into the size, shape and topology optimization of an elastic structure given certain boundary conditions. The size optimization is to determine only the size of the materials, and the shape optimization refers to the outer shape of materials. The purpose of topology optimization is to find the optimal layout of a structure included the holes within a specified region. The only known quantities in the problem are the applied loads, the possible support conditions, the volume of the structure to be constructed and possibly some additional design restrictions such as the location and size of prescribed holes or solid areas. In this problem the physical size and the shape, and the connectivity of the structure are unknown. The area of the topology optimization of continuum structure is now dominated by methods that employ the material distribution concept, like the SIMP (Bendsøe and Sigmund, 2003) and BESO (Huang and Xie, 2010).

The topology optimization problem refers to the saddle-point variation one as well as the so-called Stokes flow one of incompressive fluid (Brezzi and Fortin, 1991). The checkerboard pattern often appear in the results computed by the SIMP and BESO in which the $Q_1$-$P_0$ element is used for FEM (Finite Element Method) as well as the Stokes problem, since these patterns are more favourable than the uniform density regions (Diaz and Sigmund, 1995, Jog and Haber, 1996, Sigmund and Petersson, 1998). Computational experiments of the SIMP and BESO have shown that the filtering of sensitivity information of the optimization problem is a highly efficient way that the checkerboard pattern disappeared. Since the filter is usually used as low-pass filter for noise cleaning in the field of the image processing, this type of filter was introduced to the problems of the topology optimization by Sigmund (1994). The function of the filter is explained as the average of the density by the density of neighbourhood, or the convolution operator of the low-pass filter (Sigmund and Petersson, 1998, Hassani and Hinston, 1999 ) . However, the theoretical basis for the filtering method is not yet understood completely. In this paper, we discuss the theoretical basis for the filtering method of SIMP and BESO and as the result, the filtering method can be explained by the theorem of partition of unity and the convolution. The paper is consisted of theory of FEM for elasticity equation of the isotropic material and of checkerboard problem (Sec.1), the calculated procedure of SIMP and BESO (Sec.2), the basic theory for the filtering (Sec.3), the computed Results of SIMP and BESO (Sec.4), and conclusion remarks.

# 1. Theory of FEM for elasticity equation of a isotropic material

Let's consider the 2-dimensional elasticity problem. The $\Omega$ is domain as shown in Figure 1.

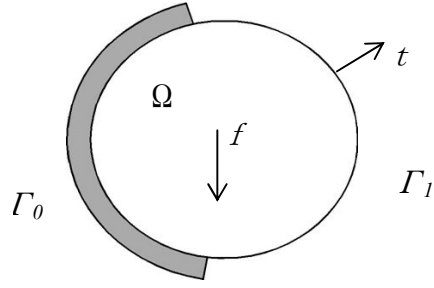

Let then $\Gamma_0$ be a part of $\Gamma$ on which we assume u = 0. We also assume the existence t in $\Omega$ of a distributed force into $f$ (eg. Gravity) and on $\Gamma_1$ of a traction $t$ that is decomposed into normal part $t_n$ and a tangential $t_t$. The governing equations are given as,

$$
\begin{aligned}
&-\partial_j \sigma_{ij}(u) = f && in\ \Omega \\
&u\big|_{\Gamma_0} = 0,\ \sigma_{nn} = 0 && on\ \Gamma_0 \\
&u\big|_{\Gamma_1} = t_n,\ t_t && on\ \Gamma_1 \\
&\sigma_{ij}(u) = 2\mu \varepsilon_{ij}(u) + \varsigma \varepsilon_{kk}(u)\delta_{ij}
\end{aligned}
\tag{1}
$$

where $\sigma_{ij}(u)$ is stress, u displacement, $f$ (eg. Gravity). the body force, t$_n$, tt and $\varepsilon_{ij}(u)$ defined as

$$
\varepsilon_{ij}(u) = \frac{1}{2}\left(\frac{\partial u_i}{\partial x_j} + \frac{\partial u_j}{\partial x_i}\right).
\tag{2}
$$

The $\mu$ and $\varsigma$ are Lame constants. The weak form of Eq.(1) is given as,

$$
2\mu \int_\Omega \varepsilon(u) : \varepsilon(v)\,dx + \varsigma \int_\Omega div\,u \cdot div\,v\,dx = \int_\Omega f \cdot v\,dx + \int_{\Gamma_1} t_n \cdot vn\,ds + \int_{\Gamma_1} t_t \cdot vt\,ds .
\tag{3}
$$

Now,

$$
\begin{aligned}
&a(u,v) = 2\mu \int_\Omega \varepsilon(u) : \varepsilon(v)\,dx, \\
&b(u,q) = \int_\Omega div\,u\ q\,dx, \\
&v \in V = (H_0^1(\Omega))^2,\ q \in Q = L^2(\Omega)
\end{aligned}
\tag{4}
$$

When $p = \varsigma\,div\,u$、 Eq.(3) is rewritten as,

$$
\begin{aligned}
&a(u,v) + b(v,p) = (f,v), && \forall v \in V, \\
&b(u,q) = \frac{1}{\varsigma}(p,q), && \forall q \in Q.
\end{aligned}
\tag{5}
$$

The existence of the solution of Eq. (5) is valid by the theorems of Lax-Milgram and the first inequality of Korn as,

$$\int_\Omega |\varepsilon(v)|^2 \, dx \geq C \|v\|^2_{1,\Omega}, \quad \forall v \in V = (\mathrm{H}^1_0(\Omega))^2. \tag{6}$$

The inf-sup condition, however, is not valid in the FEM using the $Q_1$-$P_0$ element. As a result, $div \, u = 0$ at the corner points of the macroelement formed of four quadrilaterals. When $div \, u = 0$, the material is incompressible. If the change of volume of the material like a steel generally occur, the situation is not practical. When the u is the velocity field, and $\varepsilon_{ij}(u)$ is the rate of strain, Eq. (5) is the expression of Stokes' problem for the low velocity of fluid flow. Hence, checkerboard problem of the elasticity problem occurs for the same reason as the Stokes problem why appearance of the checkerboard pattern is given in the book by Giranlt and Raviart (1986).

## 2. Calculation procedure of SIMP and BESO

We use the SIMP and BESO methods in this paper. In each of methods, we fix the design domain D consisted of both the material region $\Omega$ and the void $D\backslash\Omega$, and the single equation obtained by the virtual work principal with the Fuck's low is used. The equation is given as,

$$\mathrm{KU} = \mathrm{F}, \tag{7}$$

where U and F are the global displacement and force vectors, respectively. K is the global stiffness matrix, $u_i$ and $k_i = E(\rho_i)k_0$ are element displacement vectors and stiffness matrix, respectively.

First, the design domain is assumed to be rectangular and discretised by square finite elements. In the SIMP codes (Sigmund, 2001, Andreassen et al., 2011), a topology optimization problem based on the power-law is followed, i.e. each element $i$ is assigned a density $\rho_i$ that determines its Young's modulus $E_i$:

$$E(\rho_i) = E_{\min} + \rho_i^p (E_0 - E_{\min}) \tag{8}$$

where $E_o$ is the Young's modulus of the material, $E_{min}$ is a very small Young's modulus assigned to void regions in order to prevent the stiffness matrix from becoming singular, and $p$ is a penalization factor (typically $p$=3) introduced to ensure black-and-white solutions.

The objective is to minimize compliance can be rewritten as:

$$\left.\begin{aligned} \min_x \quad & c(\rho) = \mathrm{U}^\mathsf{T}\mathrm{KU} = \sum_{i=1}^N E(\rho_i) u_i^\mathsf{T} k_0 u_i, \\ \text{subject to} \quad & : \frac{V(x)}{V_0} = f, \\ & : \mathrm{KU} = \mathrm{F}, \\ & : 0 < \rho_{\min} < \rho_i \leq 1, \end{aligned}\right\} \tag{9}$$

where K is the global stiffness matrix, $u_i$ and $k_i = E(\rho_i)k_0$ are element displacement vectors and stiffness matrix, respectively, x the vector of design variables, $\rho_{\min}$ a vector of minimum relative densities, $N$ the total number of elements in the design domain, $p$ the penalization power, $V(x)$ and $V_0$ are the material volume and design volume, respectively, and $f$ the prescribed volume fraction.

The optimization problem Eq. (1) is solved by means of a standard optimality criteria method as followed:

$$\rho_i^{now} = \begin{cases} \max(0, \rho_i - m) & if \quad \rho_i B_i^\eta \leq \max(0, \rho_i - m) \\ \min(0, \rho_i + m) & if \quad \rho_i B_i^\eta \geq \min(1, \rho_i + m) \\ \rho_i B_i^\eta & otherwise \end{cases} \tag{10}$$

Where $m$ is a positive move limit, $\eta$ (=1/2) is a numerical damping coefficient, and $B_i$ is obtained from the optimality condition as:

$$B_i = -\frac{\partial c}{\partial \rho_i} \bigg/ \lambda \frac{\partial V}{\partial \rho_i} \tag{11}$$

Where $\lambda$ is the Lagrangian multiplier. The sensitivities of the objective function $c$ and material volume $V$ with respect to the element densities $\rho_i$ are given by:

$$\frac{\partial c}{\partial \rho_i} = -p\,\rho_i^{\,p-1} u_i^{\mathrm{T}} k_0 u_i \tag{12}$$

$$\frac{\partial V}{\partial \rho_i} = 1 \tag{13}$$

Equation (13) is based on the assumption that each element has unit volume.

In order to ensure existence of solutions to the topology optimization problem and to avoid the formation of checkerboard patterns, they introduced the mesh-independency filter as:

$$\rho_i \frac{\widehat{\partial c}}{\partial \rho_i} = \frac{1}{\sum_{j=1}^{N} \widehat{H}_j} \sum_{j=1}^{N} \widehat{H}_j \rho_j \frac{\partial c}{\partial \rho_j} \tag{14}$$

where

$$\widehat{H}_i = r_{\min} - dist(i, j) \\ \left\{ j \in N \,\middle|\, dist(i, j) \le r_{\min} \right\}, \quad i = 1, 2, \cdots, N \tag{15}$$

The operator *dist(i,j)* is defined as the distance between the centre of element $i$ and the centre of element $j$.

On the other hand, the objective of the BESO is rewritten as:

$$\begin{aligned} \min_{\rho_i \in \Omega} \quad & c(\rho) = \frac{1}{2} \mathrm{U}^{\mathrm{T}} \mathrm{KU} = \frac{1}{2} \sum_{i=1}^{N} E(\rho_i) u_i^{\mathrm{T}} k_0 u_i, \\ \text{subject to} \quad & : \frac{V(\rho)}{V_0} = f, \\ & : \mathrm{KU} = \mathrm{F}, \\ & : \rho_i = \rho_{\min} \ \ or \ \ 1. \end{aligned} \tag{16}$$

The $\rho_i$ is discrete that $\rho_i = \rho_{\min}$ or 1 and this point different from the SIMP. Then the sensitivity number is defined as,

$$\alpha_i = \begin{cases} \dfrac{1}{2} u_i^{T} k_i u_i & if \ \ \rho_i = 1 \\ 0 & if \ \ \rho_i = \rho_{\min} \end{cases} \tag{17}$$

Now, by $\widehat{\alpha}_i^{\,k}$ of the k-th iteration and $\widehat{\alpha}_i^{\,k-1}$ of k-1, $\widehat{\alpha}_i$ is defined as,

$$\overline{\alpha}_i = \frac{1}{2}\left( \widehat{\alpha}_i^{\,k} + \widehat{\alpha}_i^{\,k-1} \right), \tag{18}$$

$$\widehat{\alpha}_i = \frac{\sum_{i=1}^{N} w_i \alpha_i}{\sum_{i=1}^{N} w_i}, \tag{19}$$

where $w_i$ is the same as $\widehat{H}_i$ of the SIMP. The optimization criteria is very simple way that $\rho_i$ at the element which $\overline{\alpha}_i$ is smallest in $\Omega$ is changed from 1 to $\rho_{\min}$ and $\rho_i$ at the element which $\overline{\alpha}_i$ is largest in D/$\Omega$ is changed from $\rho_{\min}$ to 1.

### 3. The basic theory for the filter of SIMP and BESO

The inf-sup condition is not valid in the mixed FEM used the $Q_1$-$P_0$ element that is widely used in the field of engineering (Brezzi and Fortin, 1991) and the checkerboard pattern appears (Diaz and Sigmund, 1995, Jog et al. 1996, Sigmund and Petersson, 1998). They also show that the distribution parameter optimization problem is the variation of saddle point as well as the Stokes flow problem. The restriction condition of the topology optimisation problem is

$$\frac{\partial w}{\partial \rho_i} - \lambda - \alpha(1 - 2\rho_i) = 0 , \qquad (20)$$

where $w$ is the strain energy and $\alpha$ the Lagrangian multiplier. Petersson (1999) suggested that the term $\alpha(1-2\rho_i)$ comes from $\alpha \int_D \rho_i (1-\rho_i) dx$ which is correct in the sense that solution to the problem with penalty function will approach the correct ones, as one increases the penalty actor $\alpha$. The $\alpha$ is zero from this discussion and Eq. (20) becomes

$$\frac{\partial w}{\partial \rho_i} - \lambda = 0 . \qquad (21)$$

In the neighbourhood of the saddle point, the compliance $c$ is related to the strain energy $w$,

$$c = -w . \qquad (22)$$

We obtain the following equation,

$$-\frac{\partial c}{\partial \rho_i} - \lambda = 0 , \qquad (23)$$

then, we product $\rho_i$ to the both hand side of Eq. (22), we obtain

$$\rho_i^{now} = -\frac{\rho_i}{\lambda} \frac{\partial c}{\partial \rho_i} = \rho_i B_i . \qquad (24)$$

This equation is the original optimization criteria. We find that Eq. (24) is the necessary condition that gives the checkerboard pattern of $\rho_i$. The $\rho_i \dfrac{\partial \hat{c}}{\partial \rho_i}$ of Eq. (14) is effective way not to occur this pattern, and the material density locally averaged and changed from the checkerboard pattern to the uniform density region.

First, let's remember the theorem of a partition of unity as the following **proposition** ( Folland, 1999 ). If $R^n$ is a topological space and $E \subset R^n$, a partition of unity on E is a collection $\{h_\alpha\}_{\alpha \in \Lambda}$ of function in $C(R^n, [0,1])$ such that,

- each $x \in E$ has a neighbourhood on which only finitely many $h_\alpha$'s are nonzero;
- $\sum_{\alpha \in \Lambda} h_\alpha(x) = 1$ for $x \in E$.

A partition of unity $\{h_\alpha\}$ is subordinate to an open cover $\mathfrak{U}$ of E if for $\alpha$ there exists $U \in \mathfrak{U}$ with support $(h_\alpha) \subset U$. In this paper,

$$h_i = \frac{\hat{H}_i}{\sum_{i=1}^{N} \hat{H}_i} . \qquad (25)$$

Since $h_i = \hat{H}_i \Big/ \sum\limits_{i=1}^{N} \hat{H}_i$ is the continuous functions having a compact support, and $\sum\limits_{i=1}^{N} h_i = 1$ from the theorem of partition of unity and $\rho_i$ is a subset of $C_0(R^n)$, the checkerboard patterns are disappeared.

We can also interpret the filtering defined by Eq.(19). The function of convolution between a tempered distribution and a continuous function is $C_0(D)$ which is dense in $L^2(D)$ from the formula of the Fourier transform $F^{-1}[F(\rho_i) \cdot F(\hat{H}_i)]$ where $F$ is the Fourier transform and $F^{-1}$ the inverse transform.

## 3. Computed Results of SIMP and BESO

**Figure 2** shows the 2-dimensional design domain, the boundary conditions and the external load for the cantilever beam. In this paper, all results are 2-dimensional structures. The working domain is a rectangle with zero displacement boundary condition on the left side and unit vertical point load at middle of the right side. The domain size is a rectangular $80 \times 40$ mesh. **Figure 3** is the checkerboard pattern of the cantilever beam computed by SIMP. The region of the checkerboard pattern is relatively wider than that of uniform density regions, since the pattern appears in the region of high shear stress.

**Figure 4** is the resulting shape after the filtering and the penalization in SIMP. **Figure 5** shows the change of the objective compliance vs. the iteration number of cantilever beams as shown in Figs.3 and 4. The convergence is smooth and quick. From the Figure, the energy of checkerboard patterns is slightly smaller than the uniform density regions. The region of the checkerboard pattern is relatively wider than that of uniform density regions, since the pattern appear in the region of high shear stress. The uniform density regions occur due to the character that the function space $C_0(R^n)$ is connected.

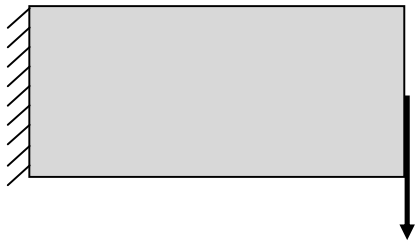

**Fig. 2** The design domain, boundary condition, and external load for the optimization of a cantilever beam

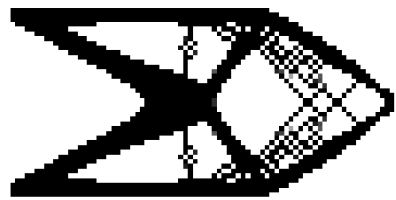

**Fig. 3** Checkerboard patterns $80 \times 40$ elements; the filter radius $r_{min} = 0.2$ with penalization without filtering, $v = 0.22$, $p = 3$, $f = 0400$

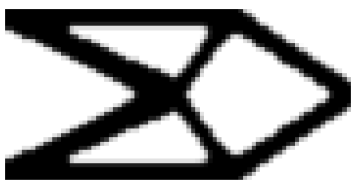

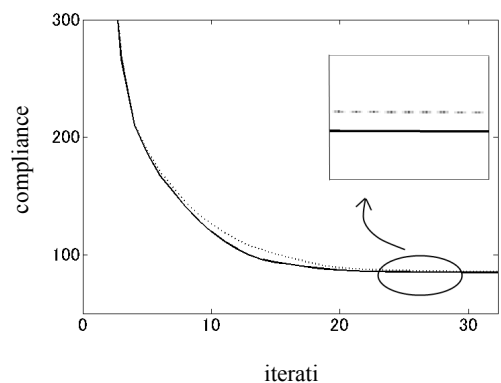

**Fig. 4** 80×40 elements; the filter radius rmin=1.3 with penalization and filtering, $\nu = 0.22$, $p = 3$, $f = 0.400$

**Fig. 5** Objective function (compliance) vs. iteration, — ; without filtering(Fig.3), --- ; with filtering (Fig.4)

**Figure 6** and **7** show the results of 40×20, 160×80 elements for the same cantilever beam, respectively. The result of 40×20 element is the same topology as that of 80×40 element and is slight different from that of 80×40. From figures, we find that this filtering method has not mesh-independency.

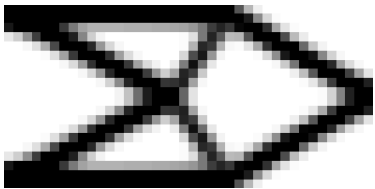

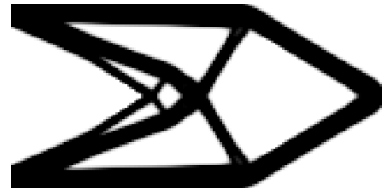

**Fig. 6** 40×20 elements; the filter radius $r_{min}$=1.5 with penalization, $\nu$ =0.22, $p = 3$, $f = 0400$

**Fig.7** 160×80 elements; the filter radius $r_{min}$=1.5 with penalization, $\nu$ =0.22, $p = 3$, $f = 0400$

**Figure 8** is the checkerboard pattern of the short cantilever beam. **Figure 9** is the resulting shape after the filtering and the penalization.

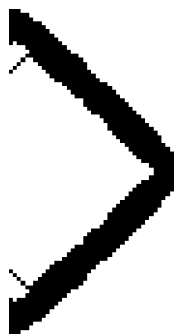

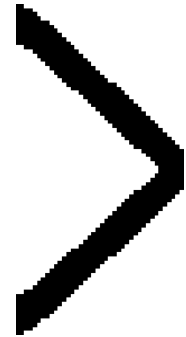

**Fig. 8** Checkerboard patterns
40×80 elements; the filter radius $r_{min}$= 0.2 with penalization without filtering, $p$ =3, $f = 025$

**Fig. 9** 40×80 elements; the filter radius rmin=1.5 with penalization and filtering, $p = 3$, $f = 0.25$

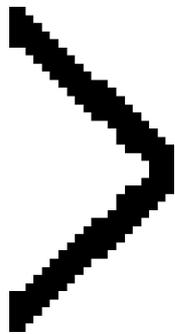

**Fig. 10**  $20 \times 40$ elements; the filter radius $r_{min}$=1.5 with penalization, $p = 3$, $f$ =0.25

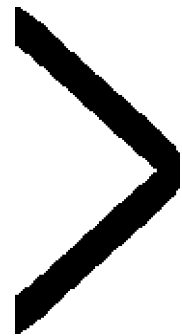

**Fig. 11**  $80 \times 160$ elements; the filter radius $r_{min}$=1.5 with penalization, $p = 3$, $f$ =025

**Figures 10 and 11** show  the results of  $20 \times 40$ and $80 \times 160$  elements for the same short cantilever beam computed by the BESO.  **Figure 12** shows that the compliances of  $40 \times 80$ and $80 \times 160$  elements vs. iteration.   From figure, we find that this filtering  method is not  mesh-independency.

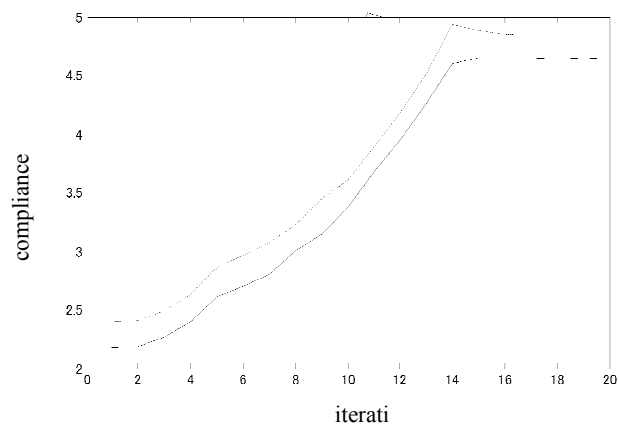

**Fig.  12**  Objective function (compliance)  vs. iteration of short cantilever
beam in BESO, $-$ ;  $80 \times 40$  elements  (Fig. 9)  --- ;  $160 \times 80$  elements  (Fig. 11)

**Conclusion remarks**

We investigate the topology optimization in which the object function is the total compliance by the SIMP and BESO. Since this problem is the saddle point variation, and the $Q_1$-$P_0$ element used for FEM is not valid for the inf-sup condition, the checkerboard pattern often appears. In order to avoid this pattern, the filter of sensitivity is used. We discuss the theoretical basis for the filtering method of SIMP and BESO. As the result, the filtering method can be explained from the theorem of partition of unity.

## Nome creature

| | | | |
|---|---|---|---|
| $c, c(\rho)$ | = | compliance | [J·m$^{-3}$] |
| $E$ | = | Young modules | [N·m$^{-2}$] |
| $F$ | = | force | [N·m$^{-3}$] |
| $f$ | = | volume ratio | [-] |
| $\hat{H}$ | = | distance defined by Eq.(15) | [-] |
| K | = | total stiffness matrix | [N·m$^{-3}$] |
| $k$ | = | stiffness matrix of element | [N·m$^{-3}$] |
| i, j | = | number of element, direction of R$^n$ | [-] |
| $p$ | = | penalty exponent | [-] |
| $r_{min}$ | = | radius of filter | [m] |
| $U$ | = | displacement vector | [m] |
| $u$ | = | displacement | [m] |
| $V(\rho)$ | = | volume of material | [m$^3$] |
| $V_0$ | = | volume of design domain | [m$^3$] |
| $w$ | = | strain energy | [J·m$^{-3}$] |

### Greek-letters

| | | | |
|---|---|---|---|
| $\varepsilon$ | = | strain | [-] |
| $\lambda$ | = | Lagrangian multiplier | [-] |
| $\rho$ | = | density | [-] |